\RequirePackage{ifpdf}
\ifpdf 
\documentclass[pdftex]{sigma}
\else
\documentclass{sigma}
\fi

\numberwithin{equation}{section}

\font\mybb=msbm10 at 12pt
\def\bb#1{\hbox{\mybb#1}}
\def\bZ {\bb{Z}}

\def\bX {\bb{X}}
\def\bY {\bb{Y}}
\def\bP {\bb{P}}

\newcommand{\Pslash}{{\bP} \hskip -.24truecm  / }

\begin{document}

\allowdisplaybreaks

\renewcommand{\thefootnote}{$\star$}

\renewcommand{\PaperNumber}{005}

\FirstPageHeading

\ShortArticleName{Quantum 3D Superparticle}

\ArticleName{The Quantum 3D Superparticle\footnote{This
paper is a contribution to the Proceedings of the Workshop ``Supersymmetric Quantum Mechanics and Spectral Design'' (July 18--30, 2010, Benasque, Spain). The full collection
is available at
\href{http://www.emis.de/journals/SIGMA/SUSYQM2010.html}{http://www.emis.de/journals/SIGMA/SUSYQM2010.html}}}

\Author{Luca MEZINCESCU~$^\dag$ and Paul K. TOWNSEND~$^\ddag$}

\AuthorNameForHeading{L.~Mezincescu and  P.K.~Townsend}

\Address{$^\dag$~Department of Physics, University of Miami, Coral Gables, FL 33124, USA}
\EmailD{\href{mailto:mezincescu@miami.edu}{mezincescu@miami.edu}}

\Address{$^\ddag$~Department of Applied Mathematics and Theoretical Physics,\\
\hphantom{$^\ddag$}~Centre for Mathematical Sciences, University of Cambridge,\\
\hphantom{$^\ddag$}~Wilberforce Road, Cambridge, CB3 0WA, UK}
\EmailD{\href{mailto:p.k.townsend@damtp.cam.ac.uk}{p.k.townsend@damtp.cam.ac.uk}}

\ArticleDates{Received November 29, 2010, in f\/inal form January 05, 2011;  Published online January 10, 2011}

\Abstract{The  minimal (${\cal N}=1$)  superparticle in three spacetime dimensions (3D)  is quantized. For  non-zero mass it describes a spin-$1/4$ semion supermultiplet of ``relativistic helicities''  $(-1/4, 1/4)$. The addition of a parity-violating Lorentz--Wess--Zumino term shifts this to $(\beta -1/4,\beta+1/4)$ for arbitrary $\beta$. For zero mass, in which case spin is not def\/ined, the quantum superparticle  describes a supermultiplet of one boson and one fermion.}


\Classification{81Q60}

\renewcommand{\thefootnote}{\arabic{footnote}}
\setcounter{footnote}{0}

\section{Introduction}

From a mathematical standpoint, an isolated  fundamental particle in a Minkowski spacetime is a unitary irreducible representation (UIR) of the  Poincar\'e group.  These UIRs are labelled by mass and some number of spin variables, depending on the spacetime dimension.  Here we shall be concerned with the three-dimensional (3D) case in which  the Lie algebra of the Poincar\'e group is spanned by a $3$-momentum ${\cal P}_\mu$ ($\mu=0,1,2)$ and a $3$-vector ${\cal J}^\mu$ that generates Lorentz transformations: the non-zero commutators are
\begin{gather*}
\left[ {\cal J}^\mu, {\cal J}^\nu\right] = i\varepsilon^{\mu\nu\rho}{\cal J}_\rho  , \qquad
\left[ {\cal J}^\mu, {\cal P}^\nu\right] = i\varepsilon^{\mu\nu\rho}{\cal P}_\rho  ,
\end{gather*}
where $\varepsilon^{012}=1$ and Lorentz indices are raised or lowered with the Minkowski metric; we use the ``mostly plus'' metric signature convention.   The UIRs of the 3D Poincar\'e group were classif\/ied by Binegar~\cite{Binegar:1981gv}. There are no spin labels for zero mass (if the possibility of continuous spin is excluded) but there are still two possible UIRs, corresponding to a distinction between bosons and fermions (see also~\cite{Deser:1991mw} for a discussion of this point). For non-zero mass, the  UIRs are classif\/ied by the two Poincar\'e invariants
\begin{gather*}
{\cal P}^2 \equiv -m^2  , \qquad {\cal P}\cdot {\cal J} = ms ,
\end{gather*}
where $m$ is the mass and $s$ is naturally thought of as ``relativistic helicity'', which can be any real number; we shall call $|s|$ the ``spin''.  In principle, $m^2$ may be negative but the UIR describes a tachyon unless $m^2\ge 0$.

Here we shall be concerned with the extension of  the Poincar\'e group to the ${\cal N}=1$  super-Poincar\'e group, the Lie superalgebra of which has an additional 2-component Majorana spinor charge ${\cal Q}^\alpha$ ($\alpha=1,2$).  The additional non-zero (anti)commutation relations are
\begin{gather*}
\left[ {\cal J}^\mu, {\cal Q}\right] = -\frac{i}{2}\Gamma^\mu {\cal Q}  , \qquad
\big\{ {\cal Q}^\alpha, {\cal Q}^\beta\big\} = \left(\Pslash C\right)^{\alpha\beta} ,
\end{gather*}
where $\Pslash = \Gamma^\mu \bP_\mu$ and $C$ is the charge conjugation matrix. We may choose $C=\Gamma^0$ in a real representation of the 3D Dirac matrices $\Gamma^\mu$,  so that the Majorana conjugate of ${\cal Q}$ is
\begin{gather*}
\bar {\cal Q} = {\cal Q}^T \Gamma^0  .
\end{gather*}
The Majorana condition is then equivalent to reality of the two spinor components of ${\cal Q}$, which become hermitian operators in the quantum theory.  When it is useful to have an explicit representation for  the 3D Dirac matrices we shall make the choice
\begin{gather*}
\Gamma^0 = i\sigma_2 , \qquad \Gamma^1 = \sigma_1  , \qquad \Gamma^2 = \sigma_3  .
\end{gather*}

To our knowledge, a systematic analysis of the UIRs of the 3D super-Poincar\'e group, along the lines of  \cite{Binegar:1981gv}, has not been undertaken but it is not dif\/f\/icult to guess what the results would be. One would expect there to be a {\it unique}  massless UIR that pairs the massless boson and fermion UIRs of the Poincar\'e group. In the massive case the UIRs must be
classif\/ied by  the two super-Poincar\'e invariants
\begin{gather*}
{\cal P}^2 \equiv -m^2   , \qquad {\cal P}\cdot {\cal J} + \frac{i}{4} \bar{\cal Q} {\cal Q} = mS  ,
\end{gather*}
where $S$ is the superhelicity of the multiplet (the average of the relativistic helicities $s$) and we expect a multiplet of
superhelicity $S$ to consist of two states of helicities $(S- 1/4,S+1/4)$   \cite{Sorokin:1992sy,Gorbunov:1997ie,Horvathy:2010vm}.
For $S=0$ this gives the $(-1/4,1/4)$ semion\footnote{Semions are particles of spin $\frac{1}{4} + \frac{n}{2}$ for integer $n$; they were initially called ``quartions''~\cite{Volkov:1989qa}; see also~\cite{Sorokin:2002st}.} supermultiplet, which has the curious feature that  the two states paired by supersymmetry have the same spin!  Since parity f\/lips the sign of helicity, this  spin-1/4 supermultiplet  also preserves parity, which just exchanges the two states. In fact, it is the  {\it unique} irreducible supermultiplet that preserves parity. This fact presumably explains why it has occurred in diverse ${\cal N}=1$ supersymmetric  3D models \cite{Witten:1999ds,Pedder:2008je,Mezincescu:2010yp}.

In this paper we shall show that all the above mentioned  supermultiplets arise from quantization of  corresponding ${\cal N}=1$ 3D superparticle models.  In general terms, a superparticle action is a super-Poincar\'e invariant action for a particle moving in a superspace extension of spacetime~\cite{Casalbuoni:1976tz,Brink:1981nb}, in  our case an ${\cal N}=1$  superspace extension  of 3D Minkowski spacetime  for which the anticommuting coordinates\footnote{The extension of classical dynamics to include anticommuting variables was f\/irst analysed systematically as long ago as 1959 \cite{Martin}.} are the two components of a Majorana spinor $\Theta$. The inf\/initesimal action of  supersymmetry on the superspace coordinates is
\begin{gather*}
\delta_\epsilon {\bX}^\mu = i  \bar\Theta \Gamma^\mu \epsilon   , \qquad \delta_\epsilon \Theta^\alpha = \epsilon^\alpha  ,
\end{gather*}
where  $\epsilon$ is a constant  anticommuting Majorana spinor parameter.  The vector space of su\-per-trans\-la\-tion-invariant dif\/ferential
1-forms on superspace is spanned by
\begin{gather*}
\Pi^\mu \equiv d{\bX}^\mu + i \bar\Theta \Gamma^\mu d\Theta  , \qquad
d\Theta^\alpha  .
\end{gather*}
The pullback of these 1-forms to the particle worldline yields manifestly super-translation invariant worldline 1-forms that may be used to construct a super-Poincar\'e invariant action. The standard massive superparticle action is
\begin{gather}\label{sparticle}
I[\bX,\bP,\Theta; \ell] =  \int  d\tau \left\{\dot {\bX}\cdot \bP  +  i \bar\Theta \Pslash \dot\Theta - \frac{1}{2}\ell \big(\bP^2 + m^2\big)\right\}  ,
\end{gather}
where $m$ is the (non-negative) mass and the overdot indicates a derivative with respect to the arbitrary worldline time parameter $\tau$.
The constraint imposed by $\ell$ is the mass-shell condition. The
constraint function  generates the inf\/initesimal gauge transformation
\begin{gather}\label{alphasym}
\delta_\alpha {\bX}^\mu = \alpha(\tau){\bP}_\mu  , \qquad \delta_\alpha {\bP}_\mu =0  , \qquad
\delta_\alpha \ell = \dot \alpha  ,
\end{gather}
with arbitrary parameter $\alpha(\tau)$. This  is  equivalent to time-reparametrization invariance (the dif\/ference is a ``trivial'' gauge transformation that vanishes as a consequence of the equations of motion).

The action \eqref{sparticle} is manifestly super-Poincar\'e invariant if we take $\bP_\mu$ (and $\ell$) to be inert under supersymmetry.
Noether's theorem then guarantees the existence of $\tau$-independent  super-Poincar\'e charges.  The Poincar\'e Noether charges are
\begin{gather}\label{Noether}
{\cal P}_\mu = \bP_\mu  , \qquad {\cal J}^\mu = \left[ \bX \wedge \bP\right]^\mu + \frac{i}{2} \bar\Theta \Theta   \bP^\mu  .
\end{gather}
The supersymmetry Noether charges are
\begin{gather}\label{N=1susy}
{\cal Q}^\alpha = \sqrt{2}  \left( \Pslash \Theta\right)^\alpha  .
\end{gather}
The action \eqref{sparticle} is also invariant under the $\bZ_2$ parity transformation
\begin{gather}\label{parity}
P: \ \ \bX^2 \to -\bX^2  , \qquad
\bP_2 \to -\bP_2  , \qquad \Theta \to \Gamma^2 \Theta  .
\end{gather}

For the massive superparticle, the superhelicity $S$ is zero classically, so one  might expect quantization to yield the semion supermultiplet with helicities $(-1/4,1/4)$. We conf\/irm this intuition in the following section, f\/irst by means of a manifestly covariant
quantization, but since the coordinates $\bX$ are non-commuting in this approach  we also present details of a variant procedure  in which the supersymplectic 2-form on the phase superspace is brought to ``super-Darboux'' form prior to quantization.  We then go on to show how the more general anyon supermultiplet with $S\ne0$ is found by adding a parity-violating Lorentz--Wess--Zumino (LWZ) term to the action, as described for the `bosonic' particle by Shonfeld \cite{Schonfeld:1980kb} (see also \cite{Cortes:1995wa}).  Quantization for $m=0$ is complicated by a  fermionic gauge invariance of the massless superparticle \cite{Siegel:1983hh},  now called ``$\kappa$-symmetry''.  To circumvent this complication,  we quantize  in the light cone gauge, thereby demonstrating that the massless 3D superparticle describes two states, one bosonic and one fermionic.

Many aspects of the results presented here are similar to those that we found in an analysis of a $\kappa$-symmetric  ${\cal N}=2$ 3D superparticle that describes, upon quantization, a centrally charged `BPS' semion supermultiplet of spins $(-1/4,-1/4,1/4,1/4)$ \cite{Mezincescu:2010gb}. That case, which we showed to have an application to Bogomolnyi vortices in the 3D Abelian-Higgs model, is similar in some respects to the
${\cal N}=1$ massive superparticle, and in other respects to the ${\cal N}=1$ massless superparticle. We promised in that work a future paper detailing results for  the general ${\cal N}=1$ superparticle, and this is it.  In the conclusions we summarize how these ${\cal N}=1$ results generalize to ${\cal N}>1$.

\section{Massive superparticle: covariant quantization}

Provided that $m\ne0$, the 3D superparticle action \eqref{sparticle} is in Hamiltonian form with
supersymplectic 2-form
\begin{gather*}
F= d\bP_\mu \Pi^\mu + i d\bar\Theta  \Pslash d\Theta =d\left[ -\bX^\mu d\bP_\mu  + i\bar\Theta \Pslash d\Theta\right]  .
\end{gather*}
The inverse of $F$ gives us the following non-zero  equal-time (anti)commutation relations:
\begin{gather}
\left[ \bX^\mu, \bP_\nu \right]   =   i\delta^\mu_\nu   , \qquad
\big\{ \Theta^\alpha ,\Theta^\beta\big\} = \frac{1}{2m^2} \left(\Pslash \Gamma^0\right)^{\alpha\beta}  , \nonumber \\
\left[ \bX^\mu , \Theta^\alpha\right]  =  \frac{i}{2m^2} \left(\Pslash \Gamma^\mu\Theta\right)^\alpha  ,
\qquad \left[ \bX^\mu , \bX^\nu \right]  = \frac{ 1}{2m^2} \bar\Theta\Theta  \varepsilon^{\mu\nu\rho} \bP_\rho   .\label{covcanon}
\end{gather}
Using these relations, one may verify that the super-Poincar\'e Noether charges have the expected (anti)commutation relations. One may also verify the {\it quantum} identities
\begin{gather}\label{quantid}
-2m^2\bar\Theta \Gamma^\mu \Theta = \bP^\mu  , \qquad \left(2im \bar\Theta\Theta \right)^2=1  .
\end{gather}
In both cases, the  right hand side is zero classically because of the classical anticommutativity of the components of $\Theta$.
Using the f\/irst of these quantum identities, one may show that
\begin{gather*}
\left[ {\cal J}^\mu, \bX^\nu \right] = i \varepsilon^{\mu\nu\rho} \bX_\rho  ,
\end{gather*}
so that $\bX$ is a 3-vector, as expected.

The second of the quantum identities \eqref{quantid} may be used to compute the relativistic helicities of the propagated modes.
From  \eqref{Noether} we see that
\begin{gather*}
m^{-1} {\cal P}\cdot {\cal J} = -\frac{1}{4}\left(2im \bar\Theta\Theta \right)   .
\end{gather*}
Next, noting that
\begin{gather*}
2im \bar\Theta\Theta = im \big[ \Theta^\alpha,\Theta^\beta\big] \left(\Gamma^0\right)_{\alpha\beta}  ,
\end{gather*}
and that the minimal  realization of the $\Theta$ anti-commutation relation is in terms  of $2\times 2$ hermitian matrices, we see that
$2im\bar\Theta\Theta$ is a traceless hermitian $2\times 2$ matrix in this realization; as it also squares to the identity,  its eigenvalues
are $1$ and $-1$,  giving us a supermultiplet of helicities $(-1/4,1/4)$.

The only dif\/f\/iculty\footnote{The ``dif\/f\/iculty'' referred to here applies equally to the massive superparticle, without WZ term, in any spacetime dimension, so it is  not a `3D peculiarity'.} with this covariant quantization is that there is no realization of the commutation relations for which the components of
$\bX$ are ``c-numbers''.  This is a consequence of the fact that the supersymplectic 2-form $F$ is not in super-Darboux form, so we now show how it can brought to this form by a change of coordinates on the phase superspace. This breaks manifest Lorentz invariance but simplif\/ies the quantization procedure, which we will show to be physically equivalent to covariant quantization.

\subsection{Quantization in super-Darboux coordinates}

We begin by writing
\begin{gather}\label{chi}
\Theta = \left(\Gamma^0\Pslash\right)^{-\frac{1}{2}} \chi  ,
\end{gather}
which def\/ines  $\chi$, a new   2-component  anticommuting variable (we do not say ``spinor''  because  the equation def\/ining $\chi$ is not covariant and hence $\chi$ does not have the Lorentz transformation properties expected of a spinor\footnote{This breaking of manifest Lorentz invariance is unavoidable: the covariant redef\/inition
{\font\mybb=msbm10 at 10pt
$\Theta \propto  \Pslash^{-\frac{1}{2}}\tilde \chi$} just leads
back to the initial Lagrangian but with $\tilde\chi$ replacing $\Theta$.}). This implies that
\begin{gather*}
d\Theta = \left(\Gamma^0\Pslash\right)^{-\frac{1}{2}}  \left[ d\chi - \frac{1}{2m} A^\mu \chi d\bP_\mu\right]   ,
\end{gather*}
where $A^\mu$ is a vector valued $2\times2$ matrix, and this gives
\begin{gather*}
i\bar\Theta \Pslash d\Theta = i\chi^T d\chi - \frac{i}{2m} \chi^T A^\mu \chi  d\bP_\mu   .
\end{gather*}
Using the identities
\begin{gather*}
\left(\Gamma^0 \Pslash\right)^{\frac{1}{2}}  \equiv  \frac{1}{\sqrt{2\big(\bP^0+m\big)}}\left(\Gamma^0 \Pslash +m\right)  ,  \nonumber \\
 \left(\Gamma^0 \Pslash\right)^{-\frac{1}{2}}  \equiv   \frac{1}{m\sqrt{2\big(\bP^0+m\big)}} \left(\Pslash \Gamma^0 +m\right)  ,
\end{gather*}
one f\/inds that
\begin{gather*}
A^0 =  \frac{1}{\big(\bP^0+m\big)} \Gamma^0\Gamma^i \bP_i  , \qquad A^i =  \Gamma^0\left(\Gamma^i +  \frac{\varepsilon^{ij} \bP_j}{\big(\bP^0+m\big)}\right)   ,
\end{gather*}
but only the antisymmetric part of $A^\mu$ contributes. We thus f\/ind that
\begin{gather*}
-\bX^\mu d\bP_\mu + i\bar\Theta \Pslash d\Theta = -\bX^0 d\bP_0 - \left(\bX^i  + \frac{i\bar\chi\chi}{2m\big(\bP^0+m\big)} \varepsilon^{ij}\bP_j\right) d\bP_i + i\chi^T d\chi   .
\end{gather*}
Let us also observe here that
\begin{gather*}
im\bar\Theta\Theta = i\bar\chi\chi  .
\end{gather*}

If we now def\/ine
\begin{gather*}
\bY^0= \bX^0   , \qquad \bY^i= \bX^i + \frac{i\bar\chi\chi}{2m\big(\bP^0+m\big)} \varepsilon^{ij}\bP_j  ,
\end{gather*}
then we f\/ind  that the  supersymplectic 2-form $F$  takes the super-Darboux form:
\begin{gather*}
F = d\bP_\mu d\bY^\mu + i d\chi^T d\chi  .
\end{gather*}
Equivalently, the superparticle action is now
\begin{gather}\label{newaction}
I[\bY,\bP,\chi; \ell] =  \int d\tau \big\{ \dot {\bY}^\mu {\bP} _\mu  + i \chi^T\dot\chi - \ell\big( {\bP}^2 + m^2\big) \big\}  ,
\end{gather}
from which we read of\/f the  new canonical (anti)commutation relations\footnote{The factor of $1/2$ in the $\chi$ anticommutator arises because each real component of $\chi$ is its own conjugate momentum. This is a second-class constraint, in Dirac's terminology, so the Poisson bracket should be replaced by a Dirac bracket prior to quantization, and this gives the factor of $1/2$.}:
\begin{gather}\label{canon}
\left[\bY^\mu,\bP_\nu\right] = i \delta^\mu_\nu    , \qquad \big\{ \chi^\alpha , \chi^\beta\big\} = \frac{1}{2} \delta_{\alpha\beta}  .
\end{gather}

When written in terms of the new variables, the super-Poincar\'e Noether charges become
\begin{gather*}
{\cal J}^0  =  \varepsilon^{ij} \bY_i \bP_j + \frac{i}{2}\bar\chi\chi  , \nonumber \\
{\cal J}^i  =  \epsilon^{ij} \left(\bY_j\bP_0- \bY_0\bP_j\right) + \frac{i\bar\chi\chi}{2\big(\bP^0+m\big)} \bP^i  ,
\end{gather*}
and
\begin{gather*}
{\cal P}_\mu = \bP_\mu  , \qquad {\cal Q}= \frac{1}{\sqrt{\big(\bP^0+m\big)}} \left(\Pslash -m \Gamma^0\right) \chi  .
\end{gather*}
The transformations of the new variables that these charges generate, using the  canonical relations \eqref{canon}, are precisely such that the super-Poincar\'e transformations of the original variables $(\bX,\Theta)$, now viewed as functions of $(\bY,\chi)$, are {\it gauge equivalent}  to the standard vector and spinor transformations generated by the Noether charges using the original  relations \eqref{covcanon}.  It follows that the Noether charges  satisfy exactly the same (anti)commutation relations whether these are computed using the covariant (anti)commutation relations  \eqref{covcanon} or with the new canonical relations \eqref{canon}, and it is easily  verif\/ied that this is indeed the case.

We have laboured the point  because of the following puzzle.  The `new' canonical relations~\eqref{canon} do {\it not} imply that the `old' phase-space variables satisfy the `old'   relations~\eqref{covcanon}. This means that the  transformation that takes the 2-form $F$ to super-Darboux form is {\it not canonical}, and this  might lead one to suspect  that  quantization in the new variables is inequivalent to the earlier covariant quantization. However, the spacetime coordinates $\bX$ are
not  physical observables because they  do not  commute with the constraint function $\bP^2+m^2$, and the same applies to $\bY$ after the redef\/inition of  coordinates.  If we restrict our attention to physical variables, i.e. those that commute ``weakly'' with the constraint function,  then we get the same results whether we use the `old' or the `new' (anti)commutation relations. In other words, the transformation of phase superspace coordinates to super-Darboux form is canonical when restricted to physical variables, and hence quantization
in the new variables, using the `new' canonical relations,  is physically equivalent to the covariant quantization using the `old', covariant,  relations.

We have still to determine the number of modes propagated by the superparticle, and their spins. For this purpose it is useful to def\/ine
the complex anticommuting variable
\begin{gather*}
\xi = \chi_1 +i \chi_2 \quad \Rightarrow \quad \big\{\xi,\xi^\dagger \big\} =1  ,
\end{gather*}
where $\chi_1$ and $\chi_2$ are, respectively, the upper and lower components of  $\chi$. The minimal matrix realization is in terms of $2\times 2$ matrices, and the  two component wave-functions may be chosen to be the eigenstates of the fermion number operator, which has
eigenvalues $0$ and $1$. This operator is
\begin{gather*}
N= \xi^\dagger \xi \quad \Rightarrow\quad  i\bar\chi\chi \to  N - \frac{1}{2}  ,
\end{gather*}
where the arrow indicates quantization using the standard operator ordering prescription for a~fermi oscillator.
We have seen how the action \eqref{newaction} is super-Poincar\'e invariant, despite appearances; it  is also still invariant under parity. Using \eqref{chi} and \eqref{parity}, we deduce that $P: \ \bY^2\to-\bY^2$ and $P: \ \bP_2 \to -\bP_2$, and
\begin{gather*}
P: \ \ \chi \to \Gamma^2 \chi  \qquad \Rightarrow \qquad P: \ \ i\bar\chi\chi \to - i\bar\chi\chi  ,
\end{gather*}
and hence
\begin{gather*}
P: \ \ N \to 1-N  .
\end{gather*}
In other words, parity exchanges the two possible eigenstates states of $N$, and hence the two states of the supermultiplet. This can be a symmetry only if these two states have the same spin, and we shall now see that this is indeed the case.

We have observed that the superhelicity $S$ is zero classically. This remains true after quantization since
\begin{gather*}
{\cal P}\cdot {\cal J} = -\frac{m}{2} \left(N-\frac{1}{2}\right)  , \qquad
\frac{i}{4} \bar{\cal Q} {\cal Q}= \frac{m}{2} \left(N-\frac{1}{2}\right)  .
\end{gather*}
The relativistic helicity operator is
\begin{gather*}
s= m^{-1} {\cal P}\cdot {\cal J} = \frac{1}{4} - \frac{N}{2}  .
\end{gather*}
For $N=0$ this gives a state of $s=1/4$ and for $N=1$ it gives a state of $s=-1/4$. We thus f\/ind the semion supermultiplet of
helicities $(-1/4,1/4)$, and hence superhelicity $S=0$.  This result depended implicitly  on the  standard operator ordering prescription for a fermi oscillator but a~dif\/ferent choice would shift the super-helicity $S$ away from zero and hence break parity (since this takes $S$ to $-S$). As the classical action preserves parity, it is natural to quantize preserving this symmetry, and it is satisfying to see that this is an automatic consequence of the usual operator ordering prescription.

\subsection{The LWZ term}

As mentioned in the introduction, it is possible to add to the massive superparticle action a~parity-violating Lorentz--Wess--Zumino
term~\cite{Schonfeld:1980kb}. This is achieved by choosing as supersymplectic 2-form
2-form
\begin{gather*}
F_\beta = F +  \frac{\beta}{2}\big({-}{\bP}^2\big)^{-\tfrac{3}{2}} \varepsilon^{\mu\nu\rho}   \bP_\mu d\bP_\nu d\bP_\rho
\end{gather*}
for arbitrary constant $\beta$.  The  $\beta$-dependent term in $F_\beta$  is manifestly super-Poincar\'e invariant and closed, but cannot be written as the exterior derivative of a manifestly Lorentz invariant 1-form. Its ef\/fect is to modify the Lorentz charges to
\begin{gather*}
{\cal J}^\mu =\ \left[ \bX \wedge \bP\right]^\mu + \left(\frac{i}{2} \bar\Theta \Theta - \frac{\beta}{m}\right) \bP^\mu  ,
\end{gather*}
so all  helicities are shifted by $\beta$, which implies that the superhelicity is now $S=\beta$.
This ef\/fect becomes particularly transparent in the light-cone gauge; the details are very similar to those presented in \cite{Mezincescu:2010gb} for the ${\cal N}=2$ centrally charged superparticle. To summarize:   the 3D massive superparticle with LWZ term describes a supermultiplet of relativistic helicities $\left(\beta -1/4, \beta+ 1/4\right)$ for arbitrary $\beta$.

A f\/inal point worth noting,  before we move on to consider the massless 3D superparticle,  is that the LWZ term is not def\/ined on mass-shell for $m=0$. This was to be expected because there are only two positive energy massless UIRs of the Poincar\'e group (excluding continuous spin irreps) and these cannot be organized into a one-parameter family of massless UIRs of the super-Poincar\'e group.

\section{Massless superparticle}

Starting from the action  \eqref{sparticle}, we set $m=0$ to get the action for a   massless 3D point particle:
\begin{gather*}
I[\bX,\bP] =  \int  d\tau \big[ \dot\bX \cdot \bP+ i\bar\Theta\Pslash \Theta - \ell \bP^2\big]  .
\end{gather*}
This action is still invariant under the ``$\alpha$-symmetry'' gauge transformations of \eqref{alphasym}.  As mentioned in the introduction, it also has a fermionic gauge invariance: the  non-zero inf\/initesimal  ``$\kappa$-symmetry'' transformations are
\begin{gather*}
\delta_\kappa \Theta =\Pslash \kappa  , \qquad \delta_\kappa {\bX}^\mu = -i \bar\Theta \Gamma^\mu \delta_\kappa \Theta  , \qquad
\delta_\kappa \ell = -4i\bar\kappa \dot\Theta  .
\end{gather*}
Although $\kappa$ has two real components, only one is relevant because $\det \Pslash=-\bP^2$, which is zero on the constraint surface; this means that $\kappa$-symmetry allows one of the two real components of $\Theta$ to be set to zero by a gauge choice, leaving one physical component.  One may verify that  not only is the action invariant under $\kappa$-symmetry but so also are the super-Poincar\'e Noether charges  of \eqref{Noether} and \eqref{N=1susy}  in the sense that their variation is zero when  $\bP^2=0$.

When $m=0$ the two-form $F$ is not invertible so the action is not in canonical form; this can also be seen from the fact that there
is  no constraint function associated with the $\kappa$-symmetry. This makes covariant quantization problematic.
Here we shall quantize in the light-cone gauge.  We f\/irst introduce  `light-cone'  coordinates and their conjugate momenta  by
\begin{gather*}
x^\pm = \frac{1}{\sqrt{2}}\big({\bX}^1 \pm {\bX}^0\big)   , \quad x = {\bX}^2    ; \qquad
p_\pm = \frac{1}{\sqrt{2}}\left({\bP}_1 \pm {\bP}_0\right)  , \quad p= {\bP}_2  ,
\end{gather*}
so that, for example, $\bP^2 = 2p_- p_+ + p^2$. We also introduce  the `light-cone' Dirac matrices
\begin{gather*}
\Gamma^\pm = \frac{1}{\sqrt{2}} \left(\Gamma^1 \pm \Gamma^0\right)  ,
\end{gather*}
so that, for example, $\bP_\mu \Gamma^\mu = p_+ \Gamma^+ + p_-\Gamma^- + p  \Gamma^2$. We now f\/ix  the $\alpha$-symmetry and
$\kappa$-symmetry invariances by setting
\begin{gather*}
x^+ =\tau  , \qquad  \Gamma^+\Theta =0  .
\end{gather*}
This implies that
\begin{gather*}
\Theta = \sqrt{\frac{1}{2\sqrt{2}\ p_-}}  \left( \begin{array}{c} \vartheta \\  0 \end{array} \right)
\end{gather*}
for some anticommuting variable $\vartheta$; the prefactor is included for later convenience. We then f\/ind that
\begin{gather*}
\Pi_\tau ^+ =1   , \qquad \Pi_\tau^- = \dot x^- + \frac{i}{2p_-}   \vartheta \dot\vartheta  , \qquad \Pi_\tau^2 = \dot x  ,
\end{gather*}
and also that $\bar\Theta\Theta =0$.  The action \eqref{sparticle} now becomes, on solving the constraint imposed by $\ell$,
\begin{gather}\label{lcgaction}
I[x,p,\vartheta] =  \int  d\tau \left\{ \dot x p + \dot x^- p_- + \frac{i}{2}\vartheta\dot\vartheta -H\right\}  , \qquad
H= \frac{p^2}{2p_-}  .
\end{gather}
The Poincar\'e charges in light-cone gauge are
\begin{gather*}
{\cal P}  =  p  , \qquad {\cal P}_- = p_-   , \qquad {\cal P}_+ = -H   , \nonumber \\
{\cal J}  =  x^- p_- + \tau H  , \qquad {\cal J}^+ = \tau p -x p_-  , \qquad {\cal J}^- =  -x^- p - xH  .
\end{gather*}
Observe that these charges are all $\vartheta$-independent. The supersymmetry charges are
\begin{gather*}
{\cal Q}^1 = \frac{p}{\sqrt{\sqrt{2}   p_-}}\ \vartheta   , \qquad {\cal Q}^2 = \sqrt{\sqrt{2}  p_-}\ \vartheta  .
\end{gather*}
Finally, parity acts as
\begin{gather*}
P: \ \  x \to -x  , \qquad p\to - p  ,
\end{gather*}
with all other variables,  including $\vartheta$,  being parity even. This is obviously a symmetry of the light-cone gauge  action \eqref{lcgaction}.

To quantize we promote the variables to operators satisfying the canonical equal-time (anti)\-com\-mu\-ta\-tion relations
\begin{gather*}
[x^-,p_-] = i  , \qquad [x,p] =i    , \qquad \left\{\vartheta , \vartheta\right\} =1  .
\end{gather*}
Using these relations, one may show that, as expected,
\begin{gather*}
\big\{ {\cal Q}^\alpha , {\cal Q}^\beta\big\} = \left( \begin{array}{cc} \sqrt{2}  H  & p \\
 p & \sqrt{2}  p_- \end{array}\right)  = \left(\Pslash \Gamma^0\right)^{\alpha\beta}  ,
\end{gather*}
and similarly that $\left[{\cal J}^\mu,{\cal Q}\right] = -\frac{i}{2}\Gamma^\mu{\cal Q}$.

The massive bosonic 3D particle was quantized in the light-cone gauge in \cite{Mezincescu:2010gb}  and it was shown there how the
commutation relations leads to a wave-function satisfying the 3D Klein--Gordon equation.  The analysis goes through for $m=0$, leading to a wave-function satisfying the wave equation, and this result carries over to the massless superparticle. The only dif\/ference is that we now have to take into account the fermionic variable $\vartheta$.  The anticommutation relation for $\vartheta$ is equivalent to $\vartheta^2=1/2$, which could be trivially realized by setting $\vartheta=1/\sqrt{2}$. However, this realization would be inconsistent with the existence of  an operator  $(-1)^F$ that anticommutes  with all fermionic operators.  In particular, we could not conclude that ${\cal Q}$ is a fermionic operator. If we insist on the existence of the  $(-1)^F$ operator then the minimal realization of  the fermion anticommutation relations is in terms of $2\times2$ (hermitian) matrices, e.g.
\begin{gather*}
\vartheta = \sigma_1/\sqrt{2}  , \qquad (-1)^F = \sigma_3  .
\end{gather*}
This gives a multiplet of two states, one a boson and the other a fermion, which is what we anticipated on the basis of the available UIRs of the Poincar\'e group.

\section{Conclusions}

In this paper we have quantized the general 3D  ${\cal N}=1$ superparticle.  We have shown that all physical  irreducible 3D  supermultiplets,  which combine the known UIRs of the 3D Poincar\'e group (excluding continuous spin irreps),  can be found in this way.

For zero mass, covariant quantization is problematic and we  used light-cone gauge quantization to show that the superparticle describes a supermultiplet of two states, one bosonic and the other fermionic.  Although spin is not def\/ined for a massless 3D particle, there is still a distinction between a boson and a fermion, which correspond to distinct unitary irreducible representations of the 3D
Poincar\'e group. It is satisfying to see that they are paired by supersymmetry.

In the case of non-zero mass, covariant quantization is mathematically straightforward  but the interpretation of results is complicated by a non-commutativity of the spacetime coordinates. This feature (which is {\it not} peculiar to 3D) is a direct result of the fact that the manifestly super-Poincar\'e supersymplectic 2-form $F$ on phase superspace is not in  super-Darboux form, and cannot be brought to this form by any {\it covariant}  change of  phase superspace coordinates.   We have shown how a {\it non-covariant}  change of coordinates can bring it to super-Darboux form, and although this change of coordinates is non-canonical in the context of arbitrary functions on the phase superspace, it is canonical if we restrict to  functions that commute with the mass-shell constraint, and it therefore yields an equivalent quantum theory.   The f\/inal result is that the massive 3D superparticle describes the semion supermultiplet of relativistic helicities $(-1/4,1/4)$. This is  the unique  irreducible ${\cal N}=1$ supermultiplet  that preserves parity, the two states of the supermultiplet being exchanged by parity. It is also an example of a supermultiplet in which all particles have the same spin.

For non-zero mass, it is possible to add to the action, consistent with all symmetries except parity, a  Lorentz--Wess--Zumino (LWZ) term. If this term has coef\/f\/icient $\beta$ then the superparticle describes the supermultiplet of relativistic helicities $(\beta-1/4,\beta +1/4)$.

In the absence of a central charge, most of the results of this paper generalize quite easily to any ${\cal N}$.  In the massless case, the single anticommuting variable $\vartheta$ that survives in the light-cone gauge will become a set $\vartheta_a$ ($a=1,\dots,{\cal N}$)
of ${\cal N}$ such variables subject to the anticommutation relation
\begin{gather*}
\left\{ \vartheta_a,\vartheta_b\right\} = \delta_{ab}  .
\end{gather*}
This will give supermultiplets with an equal number of bosons and fermions, in some representations of the universal cover of $SO({\cal N})$.  For  ${\cal N}=2$, for example,  we a get a  supermultiplet of two bosons and two fermions, both doublets of $SO(2)$. For ${\cal N}=3$ and ${\cal N}=4$ we get a supermultiplet of 4 bosons and 4 fermions. For ${\cal N}=8$  we get  a supermultiplet of 8 bosons and 8 fermions. All these supermultiplets arise  from linearization of the 11-dimensional supermembrane in various backgrounds; see e.g.~\cite{Townsend:2002wd}.

For non-zero mass, the anticommuting spinor $\chi$ that results from the f\/ield redef\/inition to super-Darboux coordinates on superspace
will become a set $\chi_a$ ($a=1,\dots,{\cal N}$) of  ${\cal N}$ such spinors. Equivalently, the one complex anticommuting variable $\xi$ will become a set $\xi_a$ of  ${\cal N}$ such variables, subject to the anticommutation relations
\begin{gather*}
\big\{\xi_a,\xi^\dagger_a\big\} = \delta_{ab}  .
\end{gather*}
In this case we get a supermultiplet of $2^{{\cal N}}$ states, with (helicity, multiplicity):
\begin{gather*}
 \left[\begin{array}{cc} -\frac{{\cal N}}{4} & 1\vspace{1mm}\\ -\frac{{\cal N}}{4} +\frac{1}{2} & {\cal N} \vspace{1mm}\\ -\frac{{\cal N}}{4} +1 &
 \frac{1}{2}{\cal N}\left({\cal N}-1\right) \vspace{1mm}\\  \vdots & \vdots \\ \frac{{\cal N}}{4} & 1
 \end{array} \right].
 \end{gather*}
The helicities increase by $1/2$ in each step and the multiplicities are the binomial coef\/f\/icients, exactly as for {\it massless} 4D
supermultiplets. For even ${\cal N}$ all states are bosons or fermions. As an example, we get a spin-2 supermultiplet for ${\cal N}=8$ that is realized by the (linearized) maximally-supersymmetric ``new massive supergravity'' \cite{Bergshoeff:2010ui}. For odd ${\cal N}$ all states are semions (i.e.\ all helicities are $1/4+ n/2$ for some integer $n$).  Adding a LWZ term will shift all the helicities by an arbitrary amount. In particular, we can choose the coef\/f\/icient of the LWZ term such that all states are bosons or fermions for odd ${\cal N}$. For ${\cal N}=7$, for example, we can arrange to get the spin-2 supermultiplet of the maximally supersymmetric linearized topologically massive gravity constructed in~\cite{Bergshoeff:2010ui}.

If central charges are allowed for  then there are many other possibilities. In particular, for ${\cal N}=2$ there is the `BPS' massive semion supermultiplet of helicities $(-\frac{1}{4},-\frac{1}{4},\frac{1}{4},\frac{1}{4})$ that we found in
\cite{Mezincescu:2010gb} from quantization of an ${\cal N}=2$ superparticle.  It is interesting to observe that the classical
superparticle model of that paper could be viewed as a massless superparticle in a~compactif\/ied 4D spacetime. In this context, it is puzzling that quantization should yield  semions  since this is not a possibility in 4D Minkowski spacetime. It may be that this puzzle is resolved by earlier work in which it was observed how semions are `almost' permitted by 4D Lorentz inva\-rian\-ce~\cite{Plyushchay:1989yx,Klishevich:2001gy}.

\subsection*{Acknowledgements}

We thank  the Benasque center for Science for a stimulating environment.  LM acknowledges partial support from National Science Foundation Award 0855386.  PKT thanks the  EPSRC for f\/inancial support and the Department of Physics at the University of Barcelona for hospitality.

\pdfbookmark[1]{References}{ref}
\LastPageEnding

\end{document}